\documentclass[a4paper]{amsart}
\usepackage{amsmath,amssymb,amsfonts}
\usepackage{amsaddr}
\usepackage{array}
\usepackage{tikz}
\usepackage{graphicx}
\usetikzlibrary{cd}

\title{An analysis of indeterminate points in discrete integrable system}
\author{Yuki Wakimoto}
\address{Department of Physics, Tokyo Metropolitan University,\\ Minami-ohsawa 1-1, Hachioji-shi, Tokyo 192-0397, Japan}
\email{wakimoto-yuki@ed.tmu.ac.jp}
\begin{document}
\maketitle
\begin{abstract}
We investigate indeterminate points in discrete integrable system. They appear in singularity confinement phenomenon naturally. We develop a method to analyse indeterminate points of dynamical maps and using this method we clarify behaviour of indeterminate points of some integrable maps. As a result, (1) we determine their indeterminacy, (2) find they are `confined' by the map, and (3) obtain a periodicity condition of the map from indeterminate point. Finally, we propose a new type of entropy using the indeterminacy in conclusion.

\end{abstract}

\section{Introduction}\label{sec:intro}
The Hirota--Miwa equation,
which governs a large class of discrete integrable systems, includes birational maps of the dimension up to infinity (instances are in \cite{doi:10.1143/JPSJ.76.024006}). This aspect has not been so far investigated in contrast to many other works devoting to clarify the structure of this equation.

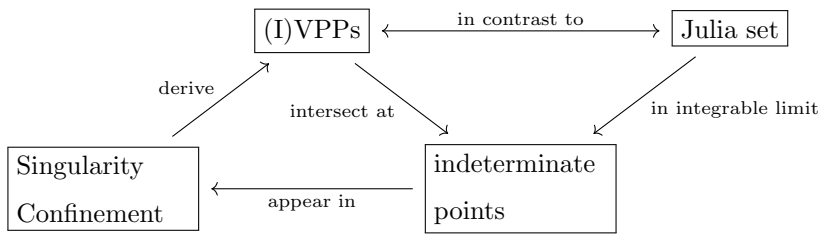
\begin{figure}[t]
\begin{equation*}
\begin{tikzcd}[row sep=large, column sep=small] &
\text{\fbox{(I)VPPs}} \ar[dr,"\text{\scriptsize intersect at}"']\ar[rr,leftrightarrow,"\text{\scriptsize in contrast to}"] && \text{\fbox{Julia set}}\ar[dl,"\text{\scriptsize in integrable limit}"] \\
\text{\fbox{\parbox[c]{6.5em}{Singularity Confinement}}} \ar[ur,"\text{\scriptsize derive}"] &
& \text{\fbox{\parbox[c]{6.5em}{indeterminate points}}}\ar[ll,"\text{\scriptsize appear in}"]
\end{tikzcd}
\end{equation*}
\caption{The schema of indeterminate points in the context of singularity confinement property, (I)VPPs, and the transition between integrable and chaotic systems}
\label{fig:schema-vsps-julia-lambda}
\end{figure}

Birational maps have indeterminate points which are mapped to $0/0$ symbolically. Moreover, they appear naturally in an extended singularity confinement sequence of integrable maps\cite{Yumibayashi:2014iha}. Singularity confinement is proposed as a test of integrability\cite{PhysRevLett.67.1825}. A counterexample of this test had been found \cite{PhysRevLett.81.325}, however, we investigated this sequence because this gives varieties of periodic points (VPP for short). VPP (more specifically, invariant VPP i.e. VPP parametrized by invariants of map, IVPP for short) appears in contrast to Julia set in chaotic systems. Julia set forms a fractal set and converges to indeterminate points in an integrable limit\cite{doi:10.1063/1.3430554}, and in this case, the indeterminate points are at the intersections of the IVPPs (their relations are shown in fig.\ref{fig:schema-vsps-julia-lambda}). Therefore we expect that the analysis of indeterminate points would contribute to the investigation of the transition between integrable and chaotic systems.

In this report, we develop a method to investigate indeterminate points (sec 2) and apply them in some archetype cases of discrete integrable birational maps with dimension 2, 3, and 6 (sec 3).

\section{Our method}
\def\tR{\tilde{\mathbb{R}}}

We show a basic idea of our method by a simple example $f(x,y)=y/x$. This function has an indeterminate point at the origin $x=0$, $y=0$ i.e. this function is invalid at this point. However, using polar coordinate $x=r\cos{\theta}$, $y=r\sin{\theta}$, this function seems to be valid at origin $r=0$; $f(r,\theta)=\tan{\theta}$. This is our basic idea. Of course, the origin is not covered by the ordinary polar coordinate. Above calculation is justified to swap the domain $\mathbb{R}^2$ of $f$ by $\tR_{(0,0)}^2$ with $\pi:\tR_{(0,0)}^2\rightarrow \mathbb{R}^2$ such that $\pi^{-1}(0,0)\simeq S^1$ and $\tR_{(0,0)}^2\setminus \pi^{-1}(0,0) \simeq \mathbb{R}\setminus \{(0,0)\}$. This $S^1$ corresponds to the region $\{(r,\theta);r=0, 0<\theta\leq 2\pi\}$ of above polar coordinate. Lifted function $\tilde{f}:\tR_{(0,0)}^2\rightarrow \mathbb{R}$ is valid at every point in $\tR_{(0,0)}^2$.
\begin{figure}[t]\label{fig:PHS}
\begin{center}
\includegraphics[width=7cm]{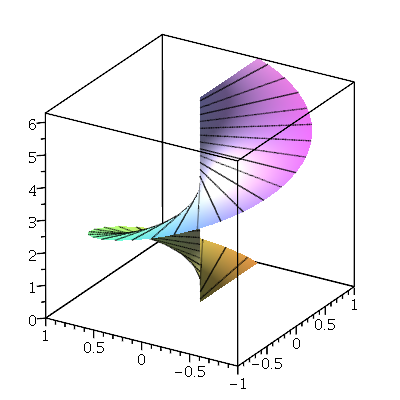}
\caption{The shape of $\tR_{(0,0)}^2$. The two end steps are identified.}
\end{center}
\end{figure}

This method is similar to the blowing-up in algebraic geometry; instead of $\tR_{(0,0)}^2$, the blowing-up prepares the space $\tR_{(0,0)}^{2\natural}=\{(x,y;t_1:t_2); xt_1-yt_2=0\}$ in $\mathbb{R}^2\times \mathbb{P}$ where $\mathbb{P}$ is a projective space $\mathbb{P}(\mathbb{R}^2)$ and consider $\pi^{\natural}:\tR_{(0,0)}^{2\natural}\rightarrow \mathbb{R}^2$. In the blowing-up case, the indeterminate point corresponds not to $S^1$, but to $\mathbb{P}$. Therefore there is a map $\natural:\tR_{(0,0)}^2 \rightarrow\tR_{(0,0)}^{2\natural}$ such that $S^1\rightarrow S^1/\mathbb{Z}_2\simeq \mathbb{P}$ and $\tR_{(0,0)}^2\setminus S^1\simeq \tR_{(0,0)}^{2\natural}\setminus \mathbb{P}$ (now $S^1=\pi^{-1}(0,0)$ and $\mathbb{P}=\pi^{\natural-1}(0,0)$). $S^1/\mathbb{Z}_2$ is obtained by an equivalence class $\theta\sim \theta+\pi$ and $S^1/\mathbb{Z}_2\simeq \mathbb{P}$ consists by $\tan\theta=t_1/t_2$.

Our method has some advantages although it might seem a variation of traditional one;
\begin{enumerate}
\item It is achieved by just introducing a coordinate which avoids indeterminate points. We can analyze a function by only one coordinate in contrast to the blowing-up which needs two coordinates for one indeterminate point.
\item It is easy to apply to general dimensional systems.
\item We can translate the result by our method to the result by traditional one by the map $\natural$.
\end{enumerate}

Some non-trivial examples are shown in next section.

\section{Indeterminate points in Integrable birational maps}
\subsection{2 dimensional M\"obius map}
\label{ssec:2dMob}
\def\2dMob{}
The first example of integrable birational map is 2 dimensional M\"obius map,
\begin{equation}
f_{\2dMob}: (x,y)\mapsto \left(x\frac{1-y}{1-x},\,y\frac{1-x}{1-y}\right),
\end{equation}
which, however, does not relate to the Hirota--Miwa equation. This map has one invariant $(x,y)\mapsto r=xy$ and reduces to ordinary M\"obius map;
\begin{equation}
\left(
\begin{matrix}
1 & -r \\
-1 & 1
\end{matrix}
\right)\in PGL(2),
\end{equation}
so this map is integrable in terms of diagonizable matrix. The inverse map is;
\begin{equation}
f_{\2dMob}^{-1}: (x,y)\mapsto \left(x\frac{1+y}{1+x},\,y\frac{1+x}{1+y}\right),
\end{equation}
therefore this map is birational. This system has two indeterminates;
$(1,1)$ and $(-1,-1)$. Let them be $\Lambda^+$ and $\Lambda^-$ respectively. Symbolically, $f_{\2dMob}(\Lambda^+)=f_{\2dMob}^{-1}(\Lambda^-)=0/0$.

To investigate these indeterminate points, we can use a bipolar coordinate whose focuses are at $\Lambda^+$ and $\Lambda^-$;
\begin{equation}
x_{\xi,\eta}=\frac{\sinh{\xi}+\sin{\eta}}{\cosh{\xi}+\cos{\eta}},\,
y_{\xi,\eta}=\frac{\sinh{\xi}-\sin{\eta}}{\cosh{\xi}+\cos{\eta}}.
\end{equation}
$\Lambda^+$ is at $\xi\rightarrow\infty$ and $\Lambda^-$ is at $\xi\rightarrow -\infty$. To consider $\xi=\infty$ and $-\infty$ with $0<\eta\leq 2\pi$, we can construct the space on which $f_{\2dMob}$ is valid. Let us denote this space by $\tR_{\Lambda^+,\Lambda^-}$.
We also denote $\pi^{-1}(\Lambda^\pm)=C^\pm$ (i.e. $C^\pm=\{(\xi,\eta);\xi=\pm\infty,0\leq \eta<2\pi\}$) where $\pi:\tR^2_{\Lambda^+,\Lambda^-}\rightarrow \mathbb{R}^2$.

$\tilde{f}_{\2dMob}(C^+)$ and $\tilde{f}_{\2dMob}^{-1}(C^-)$ can be calculated as follows; 
\begin{equation}
\tilde{f}_{\2dMob}(C^+)\ni \tilde{f}_{\2dMob}(x_{\xi,\eta},y_{\xi,\eta})\Bigl|_{\xi=\infty}=\left(\frac{\cos{\eta}-\sin{\eta}}{\cos{\eta}+\sin{\eta}},\,\frac{1+2\sin{\eta}\cos{\eta}}{2\cos^2{\eta}-1}\right),
\end{equation}
and,
\begin{equation}
\tilde{f}_{\2dMob}^{-1}(C^-)\ni\tilde{f}_{\2dMob}^{-1}(x_{\xi,\eta},y_{\xi,\eta})\Bigl|_{\xi=-\infty}=\left(\frac{\cos{\eta}+\sin{\eta}}{-\cos{\eta}+\sin{\eta}},\,\frac{-\cos{\eta}+\sin{\eta}}{\cos{\eta}+\sin{\eta}}\right).
\end{equation}
They parametrize the variety $xy=-1$ (doubly). Let us denote this variety by $\Lambda^0$. We understand that `the 0/0 for $f$ is $\Lambda^0$'. More formally, $\tilde{f}_{\2dMob}\bigr|_{C^+}:C^+\rightarrow \Lambda^0$ and $\tilde{f}_{\2dMob}^{-1}\bigr|_{C^-}:C^-\rightarrow\Lambda^0$.
Hence, the orbit with indeterminates is summarized by the diagram below,
\begin{equation}\label{eq:indeterminacy-confinement-1}
\begin{tikzcd}
& & & C^+ \ar[ld,"\tilde{f}^{-1}"']\ar[rd,"\tilde{f}"]\ar[d,dotted,"\pi"] 
& & C^- \ar[ld,"\tilde{f}^{-1}"']\ar[rd,"\tilde{f}"]\ar[d,dotted,"\pi"] \\
\cdots & \Lambda^+ \ar[l] &
\Lambda^+ \ar[l,"f^{-1}"'] &
\Lambda^+ \ar[l,"f^{-1}"'] \ar[rd,"f"] &
\Lambda^0 \ar[l,"f^{-1}"'] \ar[r,"f"]  &
\Lambda^- \ar[r,"f"] \ar[ld,"f^{-1}"']&
\Lambda^- \ar[r,"f"] &
\Lambda^- \ar[r] & \cdots \\
&&&& \displaystyle \frac{0}{0} \\
\end{tikzcd}.
\end{equation}

We see that our method can analyze the indeterminacy well by means of this example.

\subsection{3 dimensional Lotka--Volterra map}
\label{ssec:3dLV}
\def\3dLV{}

Discrete Lotka--Volterra equation;
\begin{equation}
p_i^{(n+1)}\left(1-p_{i-1}^{(n+1)}\right)=p_i^{(n)}\left(1-p_{i+1}^{(n)}\right),
\end{equation}
can be derived from Hirota--Miwa equation\cite{doi:10.1143/JPSJ.64.3125,Hirota1993}. Their periodic solution $p_i=p_{i+k}$ gives birational map with the dimension $k$. The 3 periodic case, $i\in \mathbb{Z}_3$, gives the 3 dimensional Lotka--Volterra map;
\begin{equation}
f_{\3dLV}:(x,y,z)\mapsto\left(
x\frac{1-y+yz}{1-z+zx},\,
y\frac{1-z+zx}{1-x+xy},\,
z\frac{1-x+xy}{1-y+xy}
\right).
\end{equation}
The inverse map is;
\begin{equation}
f^{-1}_{\3dLV}:(x,y,z)\mapsto\left(
x\frac{1-z+yz}{1-y+xy},\,
y\frac{1-x+zx}{1-z+yz},\,
z\frac{1-y+xy}{1-x+xz}
\right).
\end{equation}
So this map is birational.
The indeterminate set for $f_{\3dLV}$ is $\bar{\Lambda}^+=\{(x,\,y,\,z);1-z+zx=0,\,1-x+yz=0,\,1-y+xy=0\}$ and for $f^{-1}$ is $\bar{\Lambda}^-\{(x,\,y,\,z);1-y+xy=0,\,1-z+yz=0,\,1-x+xz=0\}$.

The 3d Lotka--Volterra map has two invariants (i.e. the map such that $\iota\circ f=\iota$), 
\begin{equation}
\iota: \left( \begin{array}{c} x\\y\\z\end{array}\right) \mapsto \left(\begin{array}{c}xyz\\(1-x)(1-y)(1-z)\end{array}\right).
\end{equation}
Let us denote this codomain space $\mathbb{R}^2$ by $I$. $\bar{\Lambda}^+$ and $\bar{\Lambda}^-$ have the same invariant $\Lambda=\{(r,s);r=s=-1\}\in I$ (i.e. $\iota(\bar{\Lambda}^-)=\iota(\bar{\Lambda}^+)=\Lambda$) although $f$ is not defined on them.

Now we do not analyze $\bar{\Lambda}^\pm$ directly, but $\Lambda$ in $I$ through singularity confinement property. The surface $\Sigma_{-1}=\{(x,y,z);1-z+zx=0\}$ can be  parametrized by the invariants $(r,s)=\iota_{\3dLV}(x,y,z)$ as;
\begin{equation}
\sigma_{-1}(r,s)=
\left(
\frac{r-s}{r+1},\,r\frac{s+1}{r-s},\,\frac{r+1}{s+1}
\right).
\end{equation}
Singularity confinement sequence of $f$ started at $\Sigma_{-1}$ is given as follows,
\begin{equation}
\Sigma_{-1}\rightarrow (\infty,0,1)\rightarrow (1,0,\infty)\rightarrow \Sigma_2 \rightarrow \Sigma_3 \rightarrow \cdots
\end{equation}
where $\Sigma_n=\sigma_n(I)=f^{(n-1)}\circ\sigma_{-1}(I)$ are surfaces in finite area in $M$. For instance,
\begin{eqnarray}
\Sigma_2&=&\left(
\frac{r+1}{s+1},\,r\frac{s+1}{r-s},\,\frac{r-s}{r+1}
\right)\\
\Sigma_3&=&\left(
\begin{array}{{>{\displaystyle}c}}
\frac{(r+1)(r^2-rs^2-3rs-s)}{(r-s)(r^2-rs+s^2+r+s+1)},\\
\frac{r(s+1)(r^2-rs+s^2+r+s+1)}{(r^2s+3rs-s^2+r)(r+1)},\\
\frac{(r-s)(r^2s+3rs-s^2+r)}{(s+1)(r^2-rs^2-3rs-s)}
\end{array}
\right).
\end{eqnarray}
Now, $\sigma_n$ are indeterminate at $\Lambda$.
So let us consider $\tilde{I}_{\Lambda}$ and $\tilde{\sigma}_n=f^{(n-1)}\circ \tilde{\sigma}_{-1}$ to introduce polar coordinate centered at $\Lambda$;
\begin{equation}
\begin{split}
r_{\rho,\theta}&=\rho \cos\theta-1 \\
s_{\rho,\theta}&=\rho\sin\theta-1.
\end{split}
\end{equation}
Let us denote $\tilde{\Sigma}_n=\tilde{\sigma}_n(\tilde{I}_{\Lambda})$. $\tilde{\Sigma}_{2}$ is,
\begin{equation}
\tilde{\Sigma}_2=\left(
\frac{\cos \theta}{\sin \theta},\,
\frac{(\rho \cos \theta -1)\sin \theta}{\cos \theta - \sin \theta},\,
\frac{\cos \theta - \sin \theta}{\cos \theta}
\right).
\end{equation}
In this expression, there is no indeterminacy at $\rho=0$. And at $\rho=0$, $\theta$ parametrizes $\bar{\Lambda}^-$ by $\tilde{\sigma}_2$. Through the same calculations, we can see that $\theta$ parametrizes $\bar{\Lambda}^+$ by $\tilde{\sigma}_{2n-1}$ and $\bar{\Lambda}^-$ by $\tilde{\sigma}_{2n}$. This phenomenon is summarized by the diagram below,
\begin{equation}\label{eq:indeterminacy-confinement}
\begin{tikzcd}
\cdots\ar[r,"f"]&\tilde{\Sigma}_{2n-1} \ar[d,"\rho=0"] \ar[r,"f"] &\tilde{\Sigma}_{2n} \ar[d,"\rho=0"] \ar[r,"f"] & \bar{\Sigma}_{2(n+1)-1}\ar[d,"\rho=0"]\ar[r,"f"]& \bar{\Sigma}_{2(n+1)}\ar[d,"\rho=0"]\ar[r,"f"] & \cdots\\
 \cdots\ar[r,"f"]&\bar{\Lambda}^+ \ar[r,dotted,"f"]& \bar{\Lambda}^- \ar[r,"f"] &\bar{\Lambda}^+\ar[r,dotted,"f"] & \bar{\Lambda}^-\ar[r,"f"]&\cdots
\end{tikzcd}.
\end{equation}

This means that the indeterminates of $f$, $\bar{\Lambda}^+$, is `mapped' to $\bar{\Lambda}^-$ and vice versa. We can see that $f$ `confines' the indeterminates through this analysis.

\subsection{3 point Toda map}
\label{ssec:3pToda}
\def\Toda{}

Discrete Toda equation
\begin{equation}
\begin{split}
I^{n+1}_{i}V^{n+1}_{i}=I^{n}_{i+1}V^{n}_{i} & \\
V^{n+1}_{n-1}+I^{n+1}_{i}=V^{n}_{i}+I^{n}_{i} & 
\end{split}
\end{equation}
can be derived from Hirota--Miwa equation\cite{doi:10.1143/JPSJ.64.3125,Hirota1993} and has rational maps of higher dimension up to infinity also. For instance, 3-point Toda map, given by 3 periodicity condition is the following;
\begin{equation}
f_{\Toda}:\left(
\begin{array}{ccc}
x&y&z\\u&v&w
\end{array}
\right) \mapsto
\left(
\begin{array}{*3{>{\displaystyle}c}}
y\frac{zu+zx+wu}{yw+yz+vw}&z\frac{xv+xy+uv}{zu+zx+wu}&x\frac{yw+yz+vw}{xv+xy+uv} \\
u\frac{yw+yz+vw}{zu+zx+wu}&v\frac{zu+zx+wu}{xv+xy+uv}&w\frac{xv+xy+uv}{yw+yz+vw} \\
\end{array}
\right),
\end{equation}
where $(V_0,V_1,V_2)=(x,y,z)$, $(I_0,I_1,I_2)=(u,v,w)$.
This inverse map is;
\begin{equation}
f^{-1}_{\Toda}:\left(
\begin{array}{ccc}
x&y&z\\u&v&w
\end{array}
\right) \mapsto
\left(
\begin{array}{*3{>{\displaystyle}c}}
z\frac{uw+wy+xy}{uv+uz+yz}&x\frac{uv+uz+yz}{vw+vx+xz}&y\frac{vw+vx+xz}{uw+wy+xy}\\
u\frac{vw+vx+xz}{uv+uz+yz}&v\frac{uw+wy+xy}{vw+vx+xz}&w\frac{uv+uz+yz}{uw+wy+xy}
\end{array}
\right).
\end{equation}

This map has four invariants;
\begin{equation}
\iota_{\Toda}:\mathbb{R}^6\rightarrow \mathbb{R}^4;\left(
\begin{array}{ccc}
x&y&z\\u&v&w
\end{array}
\right) \mapsto
(s,t,h,g)
\end{equation}
where
\begin{equation}
\begin{split}
&s=xyz,\,t=x+y+z+u+v+w,\,h=uvw-xyz, \\
&g=xy+yz+zx+uv+vw+wu+xv+yw+zu.
\end{split}
\end{equation}

The 3-point Toda map has similar situation to the 3d Lotka--Volterra map, but more complicate and higher dimensional. This map has singularity confinement property also, but this sequence includes indeterminate points\cite{Yumibayashi:2014iha};
\begin{equation}
\begin{split}
&\Sigma_{-1}\rightarrow
\left(\begin{array}{ccc}\infty&-h/g&0\\0&h/g&\infty\end{array}\right)\rightarrow
\left(\begin{array}{ccc}0/0&0&0/0\\0&0/0&0/0\end{array}\right)\rightarrow \\
&\qquad \left(\begin{array}{ccc}0&0/0&0/0\\0&0/0&0/0\end{array}\right)\rightarrow
\left(\begin{array}{ccc}-h/g&\infty&0\\0&\infty&h/g\end{array}\right)\rightarrow
\Sigma_4\rightarrow \cdots
\end{split}
\end{equation}
where
\begin{equation}
\Sigma_{-1}=
\left(
\begin{array}{*3{>{\displaystyle}c}}
\frac{s(g^2s-h^2t+hg^2)}{h^3}&\frac{h^2g}{g^2s-h^2t}&\frac{(g^2s-h^2t)h}{(g^2s-h^2t+hg^2)g}\\
-\frac{(h+s)(g^2s-h^2t)}{h^3}&-\frac{h(g^2s-h^2t+hg^2)}{(g^2s-h^2t)g}&\frac{h^2g}{g^2s-h^2t+hg^2}
\end{array}
\right),
\end{equation}
and
\begin{equation}
\Sigma_{4}=
\left(
\begin{array}{*3{>{\displaystyle}c}}
\frac{h^2g}{g^2s-h^2t}&\frac{s(g^2s-h^2t+hg^2)}{h^3}&\frac{(g^2s-h^2t)h}{(g^2s-h^2t+hg^2)g}\\
-\frac{(h+s)(g^2s-h^2t)}{h^3}&\frac{h^2g}{g^2s-h^2t+hg^2}&-\frac{h(g^2s-h^2t+hg^2)}{(g^2s-h^2t)g}
\end{array}
\right).
\end{equation}

\begin{figure}[h]
\begin{center}
\includegraphics[width=7cm,bb=0 0 325 288]{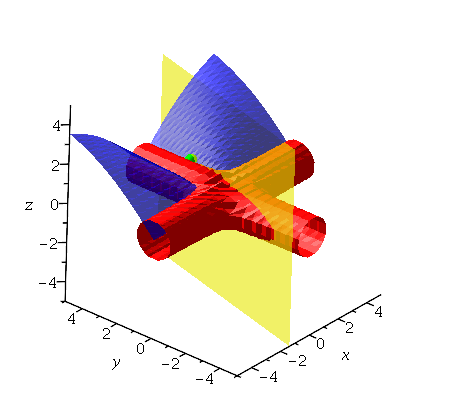}
\caption{The `cross-pipe coordinate' with $x=\rho \sin(\theta) \coth(\psi), y=\rho \sin(\theta) \cosh(\psi), z=\cos(\theta)$. The red surface is $\rho=\mathrm{const.}$ one, the blue surface is $\theta=\mathrm{const.}$ one, and yellow surface is $\psi=\mathrm{const.}$ one. The green point is specified by them.}\label{fig:crosspipe-coordinate}
\end{center}
\end{figure}

They are indeterminate at $g$-$t$ and $s$-$t$ plane (i.e. $s=h=0$ plane and $g=h=0$ one respectively). To consider them, we introduce `cross-pipe coordinate' in $t=\mathrm{const.}$ space (fig.\ref{fig:crosspipe-coordinate});
\begin{equation}\label{eq:crosspipe-coordinate}
\begin{split}
t&=t \\
s_{\rho,\theta,\psi}&=\rho \sin{\theta}\coth{\psi} \\
g_{\rho,\theta,\psi}&=\rho \sin{\theta}\cosh{\psi} \\
h_{\rho,\theta,\psi}&=\rho \cos{\theta},
\end{split}
\end{equation}
and construct the space $\tilde{I}_{\mbox{\scriptsize $g$-$t$,$s$-$t$}}$.

On this space, above sequence can be calculated as follows;
\begin{equation}
\begin{split}
&\Sigma_{-1}\rightarrow
\left(\begin{array}{*3{>{\displaystyle}c}}\infty&-\frac{h_{1,\theta,\psi}}{g_{1,\theta,\psi}}&0\\0&\frac{h_{1,\theta,\psi}}{g_{1,\theta,\psi}}&\infty\end{array}\right)\rightarrow
\left(\begin{array}{*3{>{\displaystyle}c}}\frac{1}{2}\frac{g_{\rho,\theta,\psi}}{t}&0&\infty\\0&\frac{1}{2}\frac{g_{\rho,\theta,\psi}}{t}&\infty\end{array}\right)\rightarrow \\
&\qquad \left(\begin{array}{*3{>{\displaystyle}c}}0&\frac{1}{2}\frac{g_{\rho,\theta,\psi}}{t}&\infty\\0&\infty&\frac{1}{2}\frac{g_{\rho,\theta,\psi}}{t}\end{array}\right)\rightarrow 
\left(\begin{array}{*3{>{\displaystyle}c}}-\frac{h_{1,\theta,\psi}}{g_{1,\theta,\psi}}&\infty&0\\0&\infty&\frac{h_{1,\theta,\psi}}{g_{1,\theta,\psi}}\end{array}\right)\rightarrow
\Sigma_4\rightarrow \cdots
\end{split}.
\end{equation}
The indices 1 in ${h_{1,\theta,\psi}}/{g_{1,\theta,\psi}}$ are given by $\rho$ reduction in the fraction. So this term equals $h/g$ in $\rho>0$ region.

We can check that the 3-point Toda map gives the same sequence to (\ref{eq:indeterminacy-confinement}).

Moreover, we can derive IVPP (periodic points parametrized by invariants, see \S\ref{sec:intro}) from this expression. To consider 3-periodicity condition of above sequence;
\begin{equation}
\left(\begin{array}{*3{>{\displaystyle}c}}\frac{1}{2}\frac{g_{\rho,\theta,\psi}}{t}&0&\infty\\0&\frac{1}{2}\frac{g_{\rho,\theta,\psi}}{t}&\infty\end{array}\right)
=\Sigma_4,
\end{equation}
in $(s,t,h,g)$ coordinate, $g=0$ and $t=0$ are obtained. Surely, this is general 3-periodicity condition of 3-point Toda map i.e. the third IVPP. This had been found in the previous work\cite{Yumibayashi:2014iha}, however, not completely because it contains the extra factor which corresponds to $\Lambda$. 

\section{Conclusion and Discussion}

In this report, we developed the method to analyze indeterminate points at first. It has relations to the ordinary blowing-up in algebraic geometry. The ordinary one blows up a point by $\mathbb{P}$. It conserves continuity of a space at the point, but needs two coordinates. In contrast, our method blows up by $S^1$. It makes a hole in the space, but keeps to require only one coordinate. Therefore it is easy for this method to apply to the higher dimensional case. It reduces to ordinary blowing up by double covering $S^1\rightarrow \mathbb{P}$.

We analyzed the indeterminate points of some integrable birational maps to use this method. As the result, we showed
\begin{enumerate}
\item Indeterminate points of the map and its inversion are confined by two different ways in our archetypical cases. The first way is achieved by (\ref{eq:indeterminacy-confinement-1}) in 2 dimensional M\"obius map (\S\ref{ssec:2dMob}) and the second by (\ref{eq:indeterminacy-confinement}) in 3 dimensional Lotka--Volterra map and in 3-point Toda map with dimension 6 (\S\ref{ssec:3dLV} and \S\ref{ssec:3pToda}). \label{enum:conc-1}
\item In \S\ref{ssec:3pToda}, it is determined that the indeterminacy $0/0$ in the singularity confinement sequence of 3-point Toda map. And IVPP without extra factor is found directly from this indeterminate point.
\end{enumerate}
We would like to emphasize here that the property \ref{enum:conc-1} is broken by some variations even though the birationarity and singularity confinement property is maintained. For example, a variation of the 2 dimensional M\"obius map;
\begin{equation}
(x,y)\mapsto\left(x\frac{1-y}{1-x}+a,y\frac{1-x}{1-y}\right),
\end{equation}
is still birational and has singularity confinement property. However their indeterminate points are not confined, but the map of this set gives rise to higher degree varieties at every step of the mapping. This implies we can define a new type of entropy specific to rational maps, like algebraic one, by indeterminate points. For example,
\begin{equation}
\lim_{n\rightarrow \infty} \frac{\mathrm{deg}\,f^n(\Lambda)}{n},
\end{equation}
where $f(\Lambda)=0/0$ symbolically but gives algebraic variety by the method we proposed. A deeper investigation on the relation between the behavior of the indeterminate points and the integrability of the map is left to our future work.

\section{Acknowledgement}
I would like to give thanks to Dr. S. Saito, Dr. T. Yumibayashi, and Mr. T. Takagi for stimulating discussions.

\bibliographystyle{plain}
\bibliography{reference}
\end{document}